# GazBy: Gaze-Based BERT Model to Incorporate Human Attention in Neural Information Retrieval

Sibo Dong*
sd1242@georgetown.edu
InfoSense, Dept. of Computer Science
Georgetown University, USA

Justin Goldstein
jjg130@georgetown.edu
InfoSense, Dept. of Computer Science
Georgetown University, USA

Grace Hui Yang
grace.yang@georgetown.edu
InfoSense, Dept. of Computer Science
Georgetown University, USA

## ABSTRACT

This paper is interested in investigating whether human gaze signals can be leveraged to improve state-of-the-art search engine performance and how to incorporate this new input signal marked by human attention into existing neural retrieval models. In this paper, we propose GazBy (**Gaz**e-based **B**ert model for document relevanc**y**), a light-weight joint model that integrates human gaze fixation estimation into transformer models to predict document relevance, incorporating more nuanced information about cognitive processing into information retrieval (IR). We evaluate our model on the Text Retrieval Conference (TREC) Deep Learning (DL) 2019 and 2020 Tracks. Our experiments show encouraging results and illustrate the effective and ineffective entry points for using human gaze to help with transformer-based neural retrievers. With the rise of virtual reality (VR) and augmented reality (AR), human gaze data will become more available. We hope this work serves as a first step exploring using gaze signals in modern neural search engines.

## 1 INTRODUCTION

Neural retrievers apply deep neural networks to solving information retrieval problems such as document retrieval and passage retrieval. Neural networks are given a task to score a query-document pair based on the relevance of the document to the query. These methods are usually categorized into *representation-based* (e.g., DSSM [26] and SNRM [74]) and *interaction-based* neural retrievers (e.g., DRMM [18], KRNM [10], and DeepTileBars [62]).

Representation-based neural retrievers encode a query $q$ and a document $d$ separately and interact them later when calculating a relevance score between $q$ and $d$. Each query and each document is encoded into a single fixed-size embedding vector. On the other hand, interaction-based neural retrievers generate a query-document interaction matrix in indexing-time or in query-time and feed it into neural networks to predict a relevance score between $q$ and $d$. Note that in representation-based methods, a query-document interaction appear late at the retrieval stage; whereas in interaction-based methods, the query-document interaction appears early since they take care of the $(q, d)$ pair as a whole starting from the very beginning – either from the indexing phase (e.g. when building an inverted index as in conventional retrieval methods) or from the beginning of a re-ranking process.

Currently, nearly all the most effective neural retrieval models, including MonoBERT [45] and ColBERT [33], make use of BERT [5, 6, 11, 38], a deep language model (LM) pre-trained by the transformer architecture [64] on the tasks of Masked Language Modeling and Next Sentence Prediction. MonoBERT [45] is an example of interaction-based neural retriever directly using the BERT embeddings. It inputs a '[CLS]' token appended to the front of a query-document pair as a whole input sequence to a series of encoder layers for all-in-all interactions and scores the query-document relevance based on the learned representation of the [CLS] token. On the other hand, ColBERT [33] is an example of representation-based neural retriever directly using the BERT embeddings. It creates query and document representations separately and passes them to two encoders and proposes *MaxSim*, a weighted summation of cosine similarity scores of the most similar document term to each query term, to calculate the query-document relevance. They are also known as *cross-encoder* and *bi-encoder* transformer-based models, respectively, due to the difference in their neural network architectures. In this paper, we will frequently refer to their architectures as cross-encoder and bi-encoder, respectively.[1]

BERT embeddings have also been widely used in derived functions by recent neural retrievers, mostly to obtain meaningful term-level weights. For instance, DeepCT [9] learns a linear aggregation function on BERT embeddings $a \cdot E_w(d) + b$ via regression, where $E_w(d)$ is a word $w$'s BERT embedding in document $d$. It can be thought of an aggregated contextual term weight for $w$ in $d$.

---

*Equal contribution between the first two authors. Both are first authors to the paper.

---

[1]Note that the word 'interaction' in "interaction-based retriever" and the word 'interaction' in "query-document interaction" and "late interaction" have different meanings. The former refers to a retriever's setting or architecture, where the latter refers to an actual operation between a query and a document. To avoid confusion, we choose to use cross-encoder and bi-encoder to call the two types of retrievers, instead of calling them "interaction-based" and "representation-based" as in most IR papers.

EPIC [39] used a max operation on the softplus function over BERT embeddings to select the most similar term in a document. DeepImpact [40] learned a Multi-layer Perceptron on BERT embeddings as term weight; and TILDE [78] obtained a term's log conditional probability $\log P_\theta(w|d)$ by applying a language modeling head on the [CLS] token of word $w$'s BERT embedding in $d$. Adopting the BERT representation, either directly using it or deriving from it, in neural retrievers have become a major research interest in Information Retrieval (IR).

However, the natural language processing (NLP) community reflects that looking inside of BERT's transformer blocks and layers, its attention head weights are sometimes questionable. Abnar and Zuidema [1] have shown that the BERT's attention head weights "often approximate an almost uniform distribution in higher model layers." Hollenstein and Beinborn showed that BERT's attention head weights correlate poorly with human language processing patterns [22]. Gao and Callan [16] also pointed out that "BERT directly out of pre-training has a non-optimal attention structure" and they proposed to short-circuit BERT's information flow across different transformer layers in a dense retriever. All these suggest that despite the unprecedented success these BERT-based neural retrievers have achieved, the core element that they all leverage, the vanilla BERT, is not as perfect as we think.

One recent attempt to improve the vanilla BERT embeddings is to seek help from another source of attention, *human gaze*. Human gaze modeling used to be dominated by models that implement cognitive theories with hand-crafted features (e.g., the E-Z Reader [54], Uber Reader [65], and the SWIFT Model [13]), which, although easy to explain, are often difficult to use in machine learning pipelines [61]. Modern human gaze modeling uses neural networks to predict gaze fixation durations. Popular methods include the trade-off model [20], gradient-based method [22], Recurrent Neural Networks (RNNs), and word skipping probability prediction [20, 42].

The use of human gaze can be found in computer vision [30, 59, 70, 72, 73] and natural lanauge processing (NLP) [34, 57, 61, 76] tasks, such as visual question answering [51], object referral [63], paraphrasing and sentence compression [61], sequence labeling [34], classifying pronouns [71], key phrase extraction [76], and prediction of multi-word expressions [57]. Human gaze incorporates more nuanced information about human cognitive processing because it is highly correlated with relative importance for reading comprehension [41, 53]. The use of gaze has been explored in the IR literature, too [3, 19, 35, 37, 58]. It was mostly used as a way to predict text salience and to understand relevance. However, existing work does not target pre-trained transformer models for ad hoc retrieval. To the best of our knowledge, human gaze attention has not been applied to document or passage retrieval in neural retrievers.

In this paper, we explore incorporating gaze signals into two of the most effective transformer-based neural information retrieval methods [38]. One for MonoBERT [46], and another for ColBERT [33]. With a light-weighted human gaze prediction model, we test one proof-of-concept that if we can enhance the computational attention scores in transformers using human attention scores and improve ad hoc retrieval. We propose GazBy (**Gaz**e-based **B**ert model for document relevanc**y**), a joint model that integrates human gaze fixation estimation into transformer models to predict document relevance, incorporating more nuanced information about cognitive processing into IR. Our joint model has a gaze prediction component and a relevance scoring component. We train our gaze prediction model using the GECO [4] and Zuco [24] gaze datasets and test it with both MonoBERT and ColBERT on the passage re-ranking task in the Text Retrieval Conference (TREC) Deep Learning (DL) 2019 and 2020 Tracks [5, 6].

A major effort in this work is that we investigate *how* to incorporate human gaze attention into transformer-based retrieval models. We explore extensively in this work to find out the best ways to integrate human gaze attention with computational contextual attention to help with ad hoc retrieval.

Our finding is that it is important to incorporate human gaze attention at a specific moment during the entire retrieval pipeline. Our experiments show that the **human attention should be merged into a transformer-based retrieval model when its query-document interaction happens:**

- For MonoBERT, where its interaction operation happens early during the all-in-all attention interaction, especially when the interaction is close to finish, we find it is best to merge the human gaze predictions with the last attention layer;
- For ColBERT, where its interaction operation happens late during the calculation of *MaxSim*, we find it is best to merge the human gaze predictions into the retriever during this calculation.

We also investigate other places to merge gaze in their neural network architectures, however, they produce significantly poorer results. For instance, if we incorporate the gaze attention scores into the attention mechanism in ColBERT's transformer layer, which happens before the interaction, it hurts the system's effectiveness significantly. Why does where the query-document interaction happen matter? We think that the influential weighting for each token representation by gaze fixation prediction should happen during the query and document interaction because gaze prediction itself is another form of interaction between query and document tokens. This interaction occurs through our gaze prediction model's transformer layers. By weighting the interaction between the query and document tokens through gaze-based human attention we align interaction operations among our gaze prediction model and our pre-trained transformer models, thus allowing our model to produce promising results in the cases described above.

To summarize, our paper makes the following contributions:

(1) A mechanism for incorporating human gaze attention into transformer-based neural retrieval models;
(2) A proof-of-concept of this mechanism on two state-of-the-art transformer-based neural information retrieval models that show promising results;
(3) An investigation of where and how to merge human gaze prediction in a neural information retrieval model.

## 2 RELATED WORK
### 2.1 Neural Information Retrieval
Ad hoc retrieval aims to find which are the best documents $D$ for a query $q$, structurally it is a task that involves two inputs, the query $q$ and the document $d$. It must perform an interaction, also known as Cartesian mathematically, between the two inputs and yields a single relevance score for the pair. This unavoidable interaction appear in different forms in different retrieval algorithms. We can find it as a similarity scoring function $score(q, d)$ (e.g., cosine similarity, dot product, neural matching kernels, and *MaxSim* in ColBERT) in traditional retrieval models and in the interaction-based neural retrieval methods; a joint operation between $q$ and $d$ (e.g., when building the inverted index in traditional retrieval methods, and the neural interaction in interaction-based neural retrieval methods); and the "scaled dot-product attention" [64] in the transformer layers of the transformer-based neural retrievers.

Neural Information Retrieval is the application of deep neural networks to solving information retrieval problems such as document retrieval and passage retrieval. The methods can be categorized into representation-based (e.g., DSSM [26] and SNRM [74]) and interaction-based (e.g., DRMM [18], KRNM [10], and DeepTileBars [62]).

Representation-based retrievers use a bi-encoder setting, which encode a query $q$ and a document $d$ separately and interact them later when calculating a similarity score between them. Each query and each document is encoded into a single fixed-size embedding vector. The embedding vectors can be either dense or sparse, which correspond to the recently popular dense retrievers (e.g., DPR[31], SBERT [55], Condenser [16], ICT [36], RocketQA [52], ANCE [69], and RepBERT[75]) and sparse retrievers (e.g. SparTerm [2], SPLADE [15], and EPIC [39]). They study pre-training or fine-tuning (mostly fine-tuned from the BERT embeddings [11]) methods to obtain low-dimensional encodings for query and for documents. Unlike results obtained from BoW representations, top-K results obtained from these learned representations cannot be efficiently found without any approximation [7, 29, 69]. ColBERT [33] is a type of representation-based retriever.

Interaction-based retrievers [12, 14, 18, 47–49, 56, 62, 68] generate a query-document interaction matrix in indexing-time or query-time and feed it into neural networks to predict a relevance score. Interaction-based retrievers (both pre-neural and neural) have been known for their better retrieval effectiveness. For instance, the very successful BM25 method is a sparse, interaction-based method in the pre-neural era. Early interaction-based neural retrieval models are not efficient and can only be used for re-ranking the results generated from a first-stage retriever. More recently, efforts have been made to create indices for these neural retrieval methods to save query-time calculations and support them to directly perform first-stage, full-length document retrieval [9, 74]. MonoBERT [45] uses a cross-encoder setting and is type of interaction-based retrieval. It employs an all-in-all fashion of extensive interaction among query terms and document terms.

In this paper, we find *when the interaction happens* a crucial piece of information to determine where to merge human gaze attention into transformer-based retrieval models. Query-document interaction can appear early in an algorithm, such as during indexing time when building the inverted index in traditional retrieval methods, at the early joint operation in interaction-based neural retrieval models, and during the all-to-all interaction in MonoBERT. It can also appear late in an algorithm. For instance, the last step of similarity calculation in representation-based neural retrieval methods, and the *MaxSim* operator in ColBERT.

## 2.2 Gaze and Human Attention

Our work is inspired by the use of human gaze and the improvement gained by it in computer vision [30, 59, 70, 72, 73] and natural lanauge processing (NLP) [34, 57, 61, 76] tasks. Qiao et al. used gaze as an optimization objective for visual question answering [51] and for filtering out irrelevant information, as in the case of using gaze for object referral in videos [63]. Henderson et al. demonstrated that video viewers without given an explicit task are likely trying to understand the meaning of a scene and direct their gaze accordingly [21]. Sood et al. 2020 used gaze to inform the tasks of paraphrasing and sentence compression [61]. Others in the field of NLP have also successfully used gaze for sequence labeling [34], classifying pronouns [71], key phrase extraction [76], and prediction of multi-word expressions [57].

Human gaze modeling used to be dominated by models that implement cognitive theories with hand-crafted features, which, although easy to explain, are often difficult to use in machine learning pipelines [61]. Popular models include the E-Z Reader [54], Uber Reader [65], and the SWIFT Model [13]. Modern modeling methods use neural networks to predict human sentence processing. For instance, Keller and Mahn used a Neural Attention Trade-off language model (NEAT) [20] to optimize the number of gaze fixations as a trade-off between information and one's time and energy. Hollenstein et al. have trained BERT to predict gaze in a multilingual corpus [23]. Keller [32] and Michaelov and Bergen [43] used Recurrent Neural Networks (RNNs) to predict gaze and indicated that RNN's architecture appears suited to the task. Matthies and Sogaard [42] and Hahn and Keller [20] have used neural models to predict word skipping probability, which is directly related to gaze fixation duration. These uses and modeling of gaze have been found to improve machine learning model performance on human language-related tasks compared to using models without gaze input or supervision.

## 2.3 Use of Gaze in IR and Neural IR

The use of gaze has been explored in IR literature before the deep learning era. In 2003, Salojarvi et al. [58] first studied eye movements and IR to learn whether it is possible to determine relevance from eye movements. Balatsoukas and Ruthven [3] explored the relationship between the use of correlation criteria and the level of correlation (relevant, partially relevant, irrelevant) based on a collection of eye movements, such as the number and length of fixations. Gwizdka and Zhang [19] examined correlations from the perspective of users' pupil dilation during web page visits and revisits and demonstrated the feasibility of predicting the relevance of web documents from eye-tracking data. Li et al. [37] performed extensive user study on eye-movement and user attention distribution during relevance-oriented reading comprehensions and created a prediction model to inference human attention during relevance judgments. Lagun and Agichtein segmented web pages and develop a generative model, *MICS* [35], to predict text salience. They modeled user attention based on the amount of time a user views a

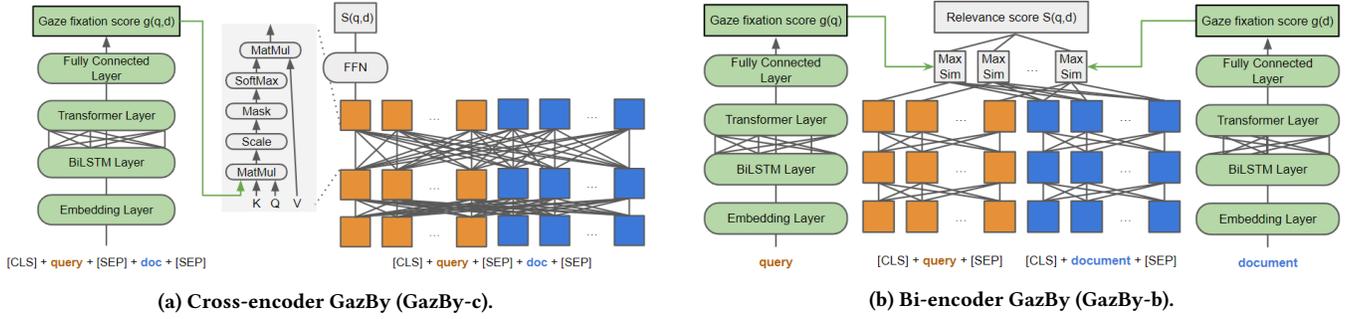

(a) Cross-encoder GazBy (GazBy-c).

(b) Bi-encoder GazBy (GazBy-b).

Figure 1: Architecture of GazBy.

web page segment on their browser along with features that describe the content of that web page segment. They demonstrated improvements on the web search, news, Wikipedia, shopping, and social network web page domains for retrieving relevant search results. Their methods, however, does not target neural information retrieval.

To the best of our knowledge, human gaze attention has not been applied to document or passage retrieval in neural retrieval models.

## 3 GAZBY: GAZE-BASED BERT MODEL FOR DOCUMENT RELEVANCY

In this paper, we propose GazBy, a light-weight joint model that combines human gaze fixation estimation into transformer-based retrieval models, with the aim to introduce more nuanced information about a text into information retrieval. We explore our idea with both MonoBERT and ColBERT, where MonoBERT is a cross-encoder (and interaction-based retriever) that performs early interactions between the query and document and ColBERT is a bi-encoder (and representation-based retriever) that performs the interaction late in its retrieval process.

### 3.1 Architecture

Figure 1 shows the overall architecture of GazBy. GazBy has two main components. The first is a gaze fixation prediction model. The second component is a transformer-based retrieval model. Depending on the architecture of the transformer-based retriever, GazBy combines the gaze prediction model differently with it. Cross-encoders are transformer-based retrieval models that create embeddings for q-d pairs using a single encoder, while bi-encoders embed q and d separately [27]. They correspond to interaction-based and representation-based neural retrievers, respectively. We therefore have two variations of GazBy: GazBy-c designed for cross-encoders (interaction-based) such as MonoBERT and GazBy-b for bi-encoders (representation-based) such as ColBERT.

The green parts of Figure 1 show the architecture of our gaze prediction model. According to [61] and [66], the combination of a BiLSTM [17] with a subsequent transformer network has the most similar predictions to human gaze processing patterns. Our gaze prediction model used in GazBy-c and GazBy-b is identical. It first takes in tokens in a pre-processed text, which could be a query or a document (as in GazBy-b), or the concatenation of a query and a document (as in GazBy-c). The tokens are fed into an embedding layer, followed by a BiLSTM layer with 128 hidden units and four self-attention transformer layers with four attention heads. Finally a fully connected feedforward network (FFN) is used to predict a gaze fixation score $g(x_i)$ for each token $x_i$ in text $\vec{x}$:

$$\vec{x} \xrightarrow{\text{BiLSTM, Transformer, FFN}} g(\vec{x}) \quad (1)$$

The higher the predicted gaze score, the longer the gaze fixation duration over a token.

Figure 1a illustrates the proposed cross-encoder GazBy (GazBy-c). It concatenates the query and the document as one input, and feeds the input into the gaze prediction model to obtain gaze fixation scores for the words. These gaze fixation scores are then used to modify a cross-encoder retrieval model's (e.g., MonoBERT) encoder layers. It is done by weighting the attention scores by the gaze scores when the encoder performs multi-head scaled dot product attention. The body of our cross-encoder adopts the BERT Large architecture, using 24 attention layers with 16 attention heads each. This joint cross-encoder model judges the relevance by the learned [CLS] token using a feed forward neural network:

$$Concat(q, d) \xrightarrow{\text{BiLSTM, Transformer, FFN}} g(q, d) \quad \text{gaze prediction}$$

$$Concat(q, d) \xrightarrow{L-1 \text{ Encoder Layers}} E_{L-1}^{(q,d)} \quad \text{input embeddings}$$

$$K_i := E_{L-1} \times W^{K_i}$$

$$V_i := E_{L-1} \times W^{V_i}$$

$$Q_i := E_{L-1} \times W^{Q_i}$$

$$E_L = Softmax(\frac{Q \times (K \odot G)^T}{\sqrt{dim}}) \times V \quad \text{modified attention}$$

$$E_L^{[CLS]} \xrightarrow{\text{FFN} S(q,d)} \quad \text{relevance score using [CLS]}$$

where $K_i, V_i, Q_i$ are $Key, Value, Query$ representations for multi-head scaled dot product attention of attention head $i$. Note this query is not our query, but an internal variable in an attention layer. In addition, $dim$ is the second dimension of $Key, Value, Query$ in each head. $G$ is $g(q, d)$ repeated over $n = q_{len} + d_{len}$ rows and $E_l$ is embeddings obtained after the $l^{th}$ encoder layer. $W^{K_i}, W^{V_i}, W^{Q_i}$ are the tensors of learned parameters that map $E_{l-1}$ to $K_i, V_i, Q_i$ $\in n \times dim$ for the $i^{th}$ attention head in the $l^{th}$ encoder layer.

**Table 1: Statistics of Gaze datasets.**

|                      | GECO [4]          | ZuCo [24]         |
| -------------------- | ----------------- | ----------------- |
| # words              | 54,364            | 20,293            |
| # sentences          | 5,300             | 1,049             |
| Avg. sentence Length | 10.64 (std=8.20)  | 19.36 (std=9.26)  |

Figure 1b is the bi-encoder (e.g., ColBERT) variant of GazBy (GazBy-b). It sends the query and the document as two separate inputs to the gaze prediction model to obtain their gaze fixation scores accordingly. It also sends the query and the document separately into a BERT Large encoder, which performs multi-head scaled dot product attention over the query or the document terms alone. Eventually, we calculate a (q,d) relevance score using a modified *MaxSim* function, which is a weighted summation of cosine similarities of the most similar document term to each query term, by weighting each similarity using its gaze fixation score:

$$q \xrightarrow{\text{BiLSTM, Transformer, FFN}} g(q) \qquad \text{gaze prediction}$$

$$d \xrightarrow{\text{BiLSTM, Transformer, FFN}} g(d)$$

$$q \xrightarrow{\text{BERT Large}} E_L^q \qquad \text{query embeddings}$$

$$d \xrightarrow{\text{BERT Large}} E_L^d \qquad \text{document embeddings}$$

$$S(q,d) = MaxSim(E_L^q \odot g(q), E_L^d \odot g(d)) \qquad \text{relevance score}$$

Next, we describe components of GazBy-b and GazBy-c in details.

## 3.2 Gaze Fixation Prediction

Gaze fixations are valuable signals from a human user that directly showcase the amount of attention the user spends while reading and processing a piece of text [53]. Given an information need, gaze fixation can indicate which words in a document and how much they capture a user's attention according to the need. In this work, we present a gaze fixation prediction model that aims to estimate the fixation duration as a score $g(x)$ for each token $x_i$ in a string of text $x = [x_1, x_2, ..., x_n]$. The bigger the value of $g(x_i)$, the longer the gaze fixation period at token $x_i$. [61] and [66] found that the combination of a BiLSTM with a subsequent transformer network has the most similar predictions to human gaze processing patterns, with four layers and four attention heads each specially. Below we detail our gaze model's design following their finding.

An input text sequence is first tokenized and then padded with special tokens. The Wordpiece [67] tokenizer is used to split input string $x$ into a sequence of tokens $x_1, x_2, ..., x_n$. To be compatible with the transformer-based retrieval models, we add special tokens such as [CLS] and [SEP] into the input embeddings, following [45]. [CLS] is used in classification tasks and usually at the very beginning of the text; [SEP] is used to separate two input text sequences. Some words are split into smaller subwords, e.g.,'wifi' to ['wi', '##fi'] and 'bluetooth' to ['blue', '##tooth'], which helps decompose the out-of-vocabulary words or unknown tokens and retain some partial meaning. We also pad the tokens with another special token [PAD] to keep all inputs the same length (empirically set to 10 in our implementation) before feeding them into the neural network. The input text is thus turned into [CLS], $x_1, x_2, ..., x_n$, [SEP], [PAD], ..., [PAD].

The neural network for gaze prediction starts with an embedding layer. GloVe embedding [50] is used to encode each token into a 300-dimension vector. We then feed the entire embedding into a single BiLSTM layer and pass it onto to a four-layers four-head self-attention transformer layer. Finally, we feed the resulting embeddings into a fully connected layer to get the gaze fixation prediction of each input token. Note that special tokens, such as [CLS], [SEP], and [PAD], are non-existing words and should not have any fixations on them. We thus label all the special tokens as [ZERO] in their ground truth labels during training, which prevents the gaze prediction model from scoring them and flatting the weights for meaningful words. Subwords are labeled with the same ground truth gaze fixation label as the word it is a part of.

Our gaze fixation prediction model is trained on two English eye-tracking datasets Ghent Eye Tracking Corpus (GECO) [4] and Zurich Cognitive Language Processing Corpus (ZuCo) [24, 25]. The GECO dataset collected eye-tracking data when native English speakers reading the entire novel *The Mysterious Affair at Styles*, which was presented in paragraphs to them on a screen. The ZuCo corpus [24] collected eye-tracking records from native English speakers when they read movie reviews, biographical sentences, and English Wikipedia. Table 1 shows the statistics of the two datasets. They are combined into a single collection and standardize their gaze fixation target value into the range of [0, 1]. As we can see, they are quite small, summing into a total of around 6,000 sentences. The neural network's training and optimization are done using Adam with a learning rate of 0.0001. We trained the model for 100 epochs and achieved a 10-fold cross-validation mean squared error of 0.004.

## 3.3 Cross-Encoder GazBy (GazBy-c)

This subsection demonstrates how our joint model GazBy-c incorporates human gaze fixations into MonoBERT, a state-of-the-art cross-encoder transformer-based retrieval model.

The proposed process takes the following steps. First, we tokenize each query-document pair and insert [CLS] and [SEP] tokens: [CLS] + $x_1...x_{q_{len}}$ + [SEP] + $x_1...x_{d_{len}}$ + [SEP]. We input this concatenated q-d pair into the gaze fixation prediction model.

Second, we also feed this q-d pair into $L - 1$ encoder layers out of $L$ total, where each encoder layer uses multi-head attention, adding and normalization, and a feed forward network, identical to Vaswani et al.'s encoder layers [64]. The input sequence to the first encoder layer is the Wordpiece embeddings of the input token sequence. Each subsequent encoder layer receives the output of the previous encoder layer as input. Multi-head attention represents the input embeddings of each encoder layer in a set of three tensors for each of several attention heads, a key tensor, $K_i$, a value tensor, $V_i$, and a query tensor, $Q_i$, using matrix multiplication against weight matrices $W^{K_i}, W^{V_i}$, and $W^{Q_i}$ respectively, for the $i^{th}$ attention head. The attention that occurs in the first L-1 encoder layers takes the form: $Softmax(\frac{Q \times K^T}{\sqrt{dim}}) \times V$. The "attention" scores here are formed by the matrix $Q \times K^T$, which denote the amount of attention that token $i$ pays to token $j$ within a sequence of length $n$.

Third, we perform a modified multi-head attention in the last encoder layer. It is done such that each token's attention scores to all other tokens are weighted by the predicted fixation duration of the token being attended to. In scaled dot product attention [64], this equates to multiplying the key layer by each term's fixation duration, expanded to size $dim$, the number of columns in the key layer representation of the query-document sequence. The newly proposed attention mechanism is

$$Softmax(\frac{Q \times (K \odot G)^T}{\sqrt{dim}}) \times V, \qquad (2)$$

where we denote element-wise multiplication by $\odot$ and matrix multiplication by $\times$. This scaled attention mechanism ensures that the most important terms have a greater impact on a query-document pair score. Words receiving more gaze have a greater impact because the attention matrix in the embedding space of $n \times n$ is meant to demonstrate how much each word in row $i$ attends to each word in column $j$. By element-wise multiplying the $K$ tensor by gaze $G \in n \times dim$, we scale the $i^{th}$ embedding's attention to the $j^{th}$ embedding by the $j^{th}$ embedding's gaze score when $K$ and $Q$ tensors interact in the attention tensor $Q \times K^T$.

In addition to this setting for GazBy-c, we also attempt to incorporate gaze weighted scaled dot product attention as described above to all layers in the encoder (GazBy-c All Layers), as well as by multiplying the Wordpiece embeddings input into the first layer using the gaze scores obtained in the first step above (GazBy-c First Layer). For results and further description, see Section 4.

Finally, we train the parameters of the gaze fixation prediction model and encoder layers' parameters by minimizing the cross-entropy loss on relevance labels as a binary classification problem.

$$Loss = -\sum_{k \in R^+} log(S(q,d)_k) - \sum_{k \in R^-} log(1 - S(q,d)_k), \qquad (3)$$

where $R^+$ are the relevant query and document pairs, and $R^-$ are the irrelevant pairs in our training data. $S(q,d)_k$ is the score for that (q,d) pair by GazBy-c.

### 3.4 Bi-Encoder GazBy (GazBy-b)

This subsection presents our gaze-based joint model, GazBy-b, for bi-encoder transformer-based retrieval models such as ColBERT [33]. Bi-encoder retrieval models feed the query and the document into two encoders and learn their representations separately and then the query-document interaction happens when calculating the (q,d) relevance score.

In this paper, we propose to incorporate the gaze fixation prediction score into ColBERT when the query and the document interact. First, we use the Wordpiece tokenizer to tokenize the query and document separately and obtain two term sequences: $q_1, q_2, ..., q_{q_{len}}$ and $d_1, d_2, ..., d_{d_{len}}$. GazBy-b appends [Q] token to the input Wordpiece query tokens, [CLS] + [Q] + $q_1...q_{q_{len}}$ + [SEP], following [33], and prepends a document indicator token, [D] to the document input tokens [CLS] + [D] + $d_1...d_{d_{len}}$ + [SEP]. We perform query augmentation by padding the query terms with BERT's [mask] tokens up to a predefined length of $M_q$ and do not augment the documents. Second, we feed the tokenized q and d terms into the gaze fixation prediction model to obtain the gaze prediction for each term, $g(q_1), g(q_2), ..., g(q_{q_{len}})$, and $g(d_1), g(d_2), ..., g(d_{d_{len}})$.

**Table 2: Statistics of TREC DL 2019-2020 passage retrieval.**

|  | Training | Development | 2019 Test | 2020 Test |
|---|---|---|---|---|
| # of queries | 502,939 | 6,980 | 43 | 54 |
| # of qrels | 532,761 | 7,437 | 9,260 | 11,386 |
| # of passages |  | 8,841,823 |  |  |

Third, simultaneously with the second step, we feed the tokenized query and document into two separate BERT encoders and fully connected layers to get the contextual representations for each term: $\overrightarrow{q_1}, ..., \overrightarrow{q_{q_{len}}}$ and $\overrightarrow{d_1}, ..., \overrightarrow{d_{d_{len}}}$. Fourth, we calculate the relevance score $S(q,d)$ by using a modified $MaxSim$ operator:

$$S(q,d) = \sum_{i}^{q_{len}} g(q_i) \cdot \max_{j} cosine(\overrightarrow{q_i}, \overrightarrow{d_j}) \cdot g(d_j), \qquad (4)$$

We optimize GazBy-b by minimizing the pairwise cross-entropy loss the same as in Eq. 3.

Note that the gaze fixation scores are added when the q-d interaction takes place. To test if it can be done at other places, we also have a variation, GazBy-b Last Layer, such that it multiplies the gaze prediction to the attention scores at the last encoder layer. We also test adding the gaze prediction scores at both the attention layers and the $MaxSim$ function (this setting is called GazBy-b Combined). The results are reported in Section 4.

## 4 EXPERIMENTS

This section reports our experiment results and findings for passage re-ranking on the TREC 2019 and 2020 Deep Learning (DL) Tracks.

### 4.1 Task and Datasets

The document collections used in TREC DL 2019-2020 Tracks are based on training data from MS MARCO, a dataset created by Microsoft in 2016 and adapted to ad hoc retrieval tasks in 2018 [44]. The datasets were created with the aim of improving ad hoc retrieval with training data that has sparse labels, mimicking "real-world" retrieval where the number of relevant documents through user click logs is sparse. MS MARCO is a collection of 8.8 million web passages and 1 million Bing user searches. No documents in the dataset are marked as irrelevant, and each query is associated with one or more positive passages.

In the pooling and judging process, the National Institute of Standards and Technology (NIST) chose a subset of these sampled queries for judging, based on budget constraints and with the goal of finding a sufficiently comprehensive set of relevance judgments to make the test collection reusable [5, 6]. This led to a judged test set of 43 queries in 2019 and 54 queries in 2020. The qrels file contains a four-point scale judgments from irrelevant (0) to perfect relevant (3). For metrics that binarize the judgment scale, we map passage judgment levels 2 and 3 to relevant and map document judgment levels 0 and 1 to irrelevant. We use TREC's evaluation scripts to compute the above metrics on our retrieved results.[2] Table 2 shows the statistics of the dataset used in our experiments.

Without loss of generality, our experiments focus on the passage re-ranking task. We evaluate all baselines and proposed models on the TREC DL 2019 and 2020 passage re-ranking task, for which we re-rank 1000 passages that were provided by NIST for each query.

---
[2]https://github.com/usnistgov/trec_eval.

### 4.2 Metrics

To evaluate the retrieval effectiveness, we employ TREC DL's official evaluation metrics. They include: Precision (P) [8] at rank position 10 $P = \frac{1}{Q} \sum_{i=1}^{|Q|} \frac{\text{relevant documents retrieved}}{10}$, averaged over all test queries; Normalized Discounted Cumulative Gain (nDCG) at rank 10 [28] $nDCG_{10} = \frac{DCG_{10}}{IDCG_{10}}$, which is made up of $DCG_{10} = \sum_{i=1}^{10} \frac{2^{rel_i}-1}{log_2(i+1)}$, the discounted cumulative gain, and $IDCG_{10} = \sum_{i=1}^{|REL_{10}|} \frac{rel_i}{log_2(i+1)}$, the ideal discounted cumulative gain, where $rel_i$ is the relevance of the result at position $i$ and $REL_{10}$ represents the list of relevant documents ordered by their relevance up to position 10; Mean Average Precision [8] $MAP = \frac{1}{Q} \sum_{i=1}^{|Q|} AP(q_i)$, averaged over all test queries; and Reciprocal Rank (RR) [8] $RR = \frac{1}{Q} \sum_{i=1}^{|Q|} \frac{1}{rank_i}$, the inverse of the first occurrence of the first relevant document in a list of results, averaged over all test queries.

### 4.3 Experimental Setup

*4.3.1 Baselines:*

- **BM25 [56]:** a traditional probabilistic retrieval model that is one of the top performing IR methods prior to deep learning. We use the Anserini toolkit[3] with all default settings to reproduce the BM25 experiments.
- **MonoBERT [46]:** a cross-encoder neural retrieval model based on BERT Large. Our implementation follows [46] and uses a query-document concatenated sequence [CLS] + $x_1...x_{q_{len}}$ + [SEP] + $x_1...x_{d_{len}}$ + [SEP] as input to BERT Large to obtain q-d term embeddings. The embedding of the [CLS] token is used for classification after being fed into a feed forward neural network. Following [45], we use 1024 dimensional vectors to represent query and document token embeddings as the hidden states of our transformer layers, with 512 total tokens in the query and document input sequence.
  Our transformer has 16 attention heads in each encoder layer, and uses a dropout rate of 0.1 on attention probabilities. It is the basis for many top performing ad hoc retrieval models [5, 6].
- **ColBERT [33]:** a bi-encoder neural retrieval model based on BERT that uses a late interaction mechanism to improve query-time efficiency, where the query and document embeddings are learned separately and only interact with each other by MaxSim function at the last stage of retrieval. It is a highly effective method with slightly worse performance in effectiveness than MonoBERT but runs much more efficient. Our implementation follows [33].
- **ColBERT + tf-idf:** a variant of ColBERT that uses tf-idf to weight the *MaxSim* scores for each query term: $S(q,d) = \sum_{i}^{q_{len}} idf(q_i) \cdot MaxSim(\overrightarrow{q_i}, d)$. It uses tf-idf, a popular term weighting scheme.

*4.3.2 Cross-encoder GazBy variations:*

*Methods merging gaze during q-d interaction:*

- **GazBy-c (or GazBy-c Last Layer):** the proposed variation of GazBy for cross-encoders as described in Section 3.3. We use MonoBERT [46] with BERT Large as our base model. The implementation details of BERT Large are as outlined in [46].
- **GazBy-c All Layers:** a variation of GazBy-c. It is identical to GazBy-c Last Layer, except that each encoder layer (24 total) uses gaze predictions to modify scaled dot product attention across 16 attention heads, instead of only modifying the last encoder layer.

*Methods merging gaze before q-d interaction:*

- **GazBy-c First Layer:** another variation of GazBy-c. It incorporates gaze prediction into the cross-encoder before it creates interaction between the (q,d) pair in the first encoder layer. We expand each gaze fixation prediction score generated by our gaze model to a dimension of 1024, such that each gaze fixation score is repeated 1024 times in the expanded vector, creating a $(q_{len} + d_{len}) \times 1024$ gaze fixation prediction tensor. We then use element-wise multiplication to multiply the input Wordpiece embedding vectors by this tensor. Finally, we feed these embeddings into BERT Large and score the relevance through the [CLS] token.

*4.3.3 Bi-encoder GazBy variations:*

*Methods merging gaze during q-d interaction:*

- **GazBy-b (or GazBy-b MaxSim):** the proposed variation of GazBy for bi-encoders as described in Section 3.4. Following the default settings of ColBERT, we pad the query and document to the max length of 32 and 180, respectively. We use BERT Large as the encoder ; and the implementation details are identical for the variations below.

*Methods merging gaze before q-d interaction:*

- **GazBy-b Last Layer:** a variation of GazBy-b that performs interaction between the (q,d) pair before their interaction in the *MaxSim* operator. Instead of using gaze predictions as term weights as in GazBy-b MaxSim, we element-wise multiply the gaze fixation predictions with the attention scores of the last layer of query and document encoders. This incorporation of gaze is similar to what is done in "GazBy-c Last Layer" (See Eq. 2). Then the un-modified *MaxSim* function is used to compute the relevance score using the resulting query and document embeddings. Other pre-processing and training details are the same with GazBy-b.
- **GazBy-b Combined:** We also test adding the gaze prediction scores both at the attention layers and at the *MaxSim* function. We combine GazBy-b MaxSim and GazBy-b Last Layer. That is, we element-wise multiply the gaze fixation predictions with the attention scores at the last layer of query and document encoders and then use modified *MaxSim* function to get the relevance score.

---
[3]https://github.com/castorini/anserini.

Table 3: Passage re-ranking results on TREC DL 2019 and 2020, separated by whether the model was Bi-Encoder or Cross-Encoder based. The best performing results are shown in bold for each section. Two arrows indicate that the change in performance from the baseline ColBERT (for GazBy-b variants) or MonoBERT (for GazBy-c variants) model is greater than 5%, while one arrow indicates the direction of the change was no greater than 5%.

| Method | DL 2019 | | | | DL 2020 | | | |
|---|---|---|---|---|---|---|---|---|
| | P@10 | nDCG@10 | MAP | RR | P@10 | nDCG@10 | MAP | RR |
| BM25 | 0.412 | 0.506 | 0.301 | 0.704 | 0.350 | 0.480 | 0.286 | 0.659 |
| MonoBERT | 0.614 | 0.703 | 0.433 | **0.881** | **0.558** | **0.701** | **0.466** | 0.796 |
| GazBy-c First Layer | 0.028 ↓↓ | 0.048 ↓↓ | 0.068 ↓↓ | 0.069 ↓↓ | 0.015 ↓↓ | 0.031 ↓↓ | 0.032 ↓↓ | 0.060 ↓↓ |
| GazBy-c All Layers | 0.033 ↓↓ | 0.050 ↓↓ | 0.061 ↓↓ | 0.080 ↓↓ | 0.024 ↓↓ | 0.028 ↓↓ | 0.030 ↓↓ | 0.051 ↓↓ |
| GazBy-c Last Layer | **0.621 1.1%↑** | **0.717 2.0%↑** | **0.438 1.2%↑** | 0.881 | 0.544 -2.5%↓ | 0.696 -0.7%↓ | 0.444 -4.7%↓ | **0.832 4.5%↑** |
| ColBERT | **0.619** | **0.713** | **0.447** | 0.861 | 0.533 | 0.698 | 0.460 | **0.856** |
| ColBERT tf-idf | 0.614 -0.8%↓ | 0.707 -0.8%↓ | 0.443 -0.9%↓ | 0.831 -3.5%↓ | 0.535 0.4%↑ | 0.699 0.1%↑ | **0.461 0.2%↑** | 0.854 -0.2%↓ |
| GazBy-b Last Layer | 0.505 ↓↓ | 0.579 ↓↓ | 0.332 ↓↓ | 0.759 ↓↓ | 0.374 ↓↓ | 0.484 ↓↓ | 0.279 ↓↓ | 0.587 ↓↓ |
| GazBy-b Combined | 0.302 ↓↓ | 0.487 ↓↓ | 0.255 ↓↓ | 0.559 ↓↓ | 0.321 ↓↓ | 0.391 ↓↓ | 0.239 ↓↓ | 0.502 ↓↓ |
| GazBy-b MaxSim | 0.616 -0.5%↓ | 0.698 -2.1%↓ | 0.429 -4.0%↓ | **0.878 2.0%↑** | **0.541 1.5%↑** | **0.704 0.9%↑** | 0.460 | 0.846 -1.2%↓ |

*4.3.4 Training, Validation, and Testing.* In our experiments, the training data takes the form of triples: (query, positive passage, negative passage). These triples are generated from TREC's provided training qrels. Validation triples are available in an identical format. We train all GazBy-c models and the MonoBERT baseline on 4k training triples for 4 epochs. Our model is based on the *Hugging Face* implementation of BERT for sequence classification.[4] The MonoBERT checkpoint is loaded from castorini[5] prior to fine-tuning, which has been trained on the MS MARCO document collection. For each epoch, we perform validation with a subset of 700 development triples to increase validation speed, and at prediction time we select the model with the highest validation accuracy. We train all GazBy-b models and the ColBERT baseline on the TREC DL train triples collection for 50,000 steps with batch size equals 32. All GazBy-c and GazBy-b models are trained using the Adam Optimizer with learning rate $3e^{-6}$ and $\epsilon = 1e^{-6}$. Three GeForce RTX NVIDIA GPUs are used for training.

## 4.4 Main results

Table 3 reports the main experimental results on search effectiveness for the official TREC DL 2019 and 2020 passage re-ranking tasks. From Table 3, we can see that "GazBy-c Last Layer" and "GazBy-b Maxsim" give the best performance among cross-encoder GazBy and bi-encoder GazBy variations, respectively. They also work well compared to their own baselines, MonoBERT and ColBERT. In particular, "GazBy-c Last Layer" outperforms MonoBERT on P@10, MAP, and nDCG@10 by 1-2%, while maintaining effectiveness in terms of RR on the TREC DL 2019 Dataset. On the TREC DL 2020 Dataset, "GazBy-c Last Layer" outperforms MonoBERT by 4.5% on RR, but decreases in effectiveness by 2.5% on P@10 and 4.7% on MAP, and 0.7% on nDCG@10. Comparing with ColBERT on DL 2019, our "GazBy-b MaxSim" run improves performance by 2.0% on RR, but decreases performance by 2.0% on nDCG@10, 4.1% on MAP, and 1% on P@10. For DL 2020, it improves on the ColBERT baseline by 2% on P@10 and 1.2% on nDCG@10, and decreases 1.2% on RR. Given that both MonoBERT and ColBERT are highly effective retrieval models that boosted the SOTA performance by significant margin from previous approaches, the improvements that we observe from Gazby are quite encouraging.

---
[4]https://huggingface.co/docs/transformers/model_doc/bert.
[5]https://huggingface.co/castorini/monobert-large-msmarco.

Note that MonoBERT and ColBERT's high performance gain rely on large scale pre-trained model that is extensively trained using superior computational powers. In contrast, the gaze model that we add on top of them is rather light-weight. Our gaze prediction model is only trained over 6,000 sentences, and our gaze prediction model has far fewer parameters (302,593) than BERT Large (345 million). Given the limited resources that our proposed method requires and limited gaze training data available, the amount of improvements that GazBy has achieved show a promising new direction in neural information retrieval.

## 4.5 Our Findings

Through Table 3, we realize that some settings of GazBy works poorly. For instance, we find that incorporating gaze into every attention layer, "GazBy-c All Layers" by performing the modified scaled dot product attention in Eq. 2 decreases model performance (as compared to "GazBy-c Last Layer") significantly by 96%. This decrease in performance shows that one cannot use gaze fixation scores in multiple places throughout the cross-encoder. We also use gaze fixation scores to modify the Wordpiece embeddings before the q-d interaction in "GazBy-c First Layer". This method severely decreases the performance of "GazBy-c Last Layer" by up to 97%.

For the bi-encoder GazBy, we find that except incorporating gaze fixation scores during the interaction between query and document using the modified *MaxSim* operator, other settings perform poorly. For instance, incorporating gaze before the interaction, as in the case of "GazBy-b Last Layer" for DL 2019 in Table 3 decreases performance by up to 25% on P@10 and 20% on RR. Similar degrades are observed on DL 2020. "GazBy-b Combined," which uses both Last Layer and MaxSim settings performs even worse.

These results show that even though gaze prediction can be a useful component to improve transformer-based retrieval models, they are sensitive to where the merging of the two components should happen. Based on what we observe from our experiments, the only effective merging point of the gaze model and the transformer model is when the query-document interaction happens.

## 5 CONCLUSION

In this paper, we investigate effective ways to incorporate human gaze attention into existing transformer-based neural information retrieval models that largely benefited from the recent development in the use of computational attention. Based on our experiments over the TREC DL 2019 and 2020 Tracks, a *key finding* is that the only effective merging point of human gaze attention and computational attention in the transformer models is when the query-document interaction happens in the retrieval algorithms.

We are intrigued by the experimental results that we observe. Based on our experiments, it is encouraging and effective to use gaze fixations to help with transformer-based retrieval models. In addition, how and where to combine the two have a significant impact on their joint effectiveness: It is not as simple as to say that we can add gaze fixation scores into the attention layers, which works for MonoBERT but not for ColBERT.

We did extensive experiments on models currently dictating the direction of the field. Two transformer-based retrievers, one interaction-based and one representation-based, are investigated. Our next immediate effort would be investigating our findings over other traditional and non-transformer neural retrievers, such as the recent dense retrievers and sparse retrievers.

We acknowledge that we make an assumption about our gaze data that may be incorrect. We assume our gaze data is well suited to the type of gaze humans apply to query and document text when determining whether they are relevant, as is the task in relevance scoring. This assumption was made out of necessity. Information retrieval is related to information need based attention. Such human attention are human attention patterns on a candidate text when the human subject is trying to understand whether that text is relevant to their needs. For example, a subject might use need-based attention when reading an article to determine whether it is relevant to a question about a related topic. Note that the human attention data we use to pre-train our model, in contrast, is comprehension based: subjects in the GeCo and ZuCo data sets are reading to understand their texts. Therefore, there is a difference between the type of the human attention data we use to pre-train our gaze prediction model and the type of attention an search engine users exhibit to determine query-document relevance. In light of this fact, our work seeks to be a starting point for future efforts. With the advent of AR/VR technology [60, 77], we are confident that devices and datasets that show the gaze of human subjects performing relevance scoring will be more available. In future work, we seek to improve GazBy by using such data (for instance, use gaze data for post-retrieval relevance feedback to neural retrievers), and to incorporate gaze prediction in non-transformer based models to understand where gaze can be best applied. The findings from this paper suggest that the interaction between query and document in these models would be a great starting place for such efforts.

## ACKNOWLEDGEMENT

The authors are thankful for the ACM SIGIR/ICTIR student travel grant, the Royden B. Davis Fellowship, a Provost's Undergraduate Research Presentation Award, and the U.S. National Science Foundation MAPWISELY Mid-Career Fund.